\documentclass[twocolumn,showpacs,amsmath,amssymb,prb]{revtex4-1}

\usepackage{slashbox}
\usepackage{hhline}
\usepackage{bm}

\usepackage{graphicx}
\def\<{\langle}
\def\>{\rangle}
\def\({\left(}
\def\){\right)}
\def\[{\left[}
\def\]{\right]}
\def\up{\uparrow}
\def\dn{\downarrow}
\def\s{\sigma}
\def\e{\mathrm{e}}
\def\i{\mathrm{i}}

\def\Im#1{\mathrm{Im}\left\{#1\right\}}

\def \chib {\chi_0}

\begin{document}
\title{Spin and Charge Dynamics Ruled by Antiferromagnetic Order in Iron Pnictides}

\author{E. Kaneshita$^{1}$}
\author{T. Tohyama$^{1,2}$}
\affiliation{
$^1$Yukawa Institute for Theoretical Physics, Kyoto University, Kyoto 606-8502, Japan\\
$^2$JST, Transformative Research-Project on Iron Pnictides (TRIP), Chiyoda, Tokyo 102-0075, Japan}

\date{\today}

\begin{abstract}
We examine the spin and charge excitations in antiferromagnetic iron pnictides by mean-field calculations with a random phase approximation in a five-band itinerant model.
The calculated excitation spectra reproduce well spin-wave dispersions observed in inelastic neutron scattering, with a realistic magnetic moment for CaFe$_2$As$_2$.
A particle-hole gap is found to be crucial to obtain consistent results; we predict the spin wave in LaFeAsO disappears at a lower energy than in CaFe$_2$As$_2$.
We analyze that the charge dynamics to make predictions for resonant inelastic x-ray scattering spectra.
\end{abstract}

\maketitle

\section{Introduction}
Iron pnictide superconductors have been intensely studied since the discovery of superconductivity in LaFeAsO$_{1-x}$F$_{x}$~\cite{kamihara}, and achieving critical temperatures ($T_c$) over 50 K.
Such a high $T_c$ evokes the high-$T_c$ cuprates, and the presence of the antiferromagnetic (AFM) order in the parent compounds suggests a close connection between the superconductivity and the magnetic order in iron pnictides as well.
From this similarity, one may expect to elucidate the origin of the high-$T_c$ mechanism, and the relation with the magnetic properties of these materials, in the same framework.

However, there exists one crucial difference between these systems ---iron pnictides are not a strongly correlated system such as the cuprates.
Although the parent compounds of the high $T_c$ cuprates are a Mott insulator, those of the iron pnictides are a metal.
Because of the itinerant feature in iron pnictides, the magnetic properties are not so readily understood as those in the insulating AFM systems.

The most powerful tool to observe the spin excitations is inelastic neutron scattering, and the observations in CaFe$_2$As$_2$ and BaFe$_2$As$_2$ (the so-called 122 system) have discovered the following characteristics of the spin excitation.
(I) The spin wave excitation is extended up to a high energy (not less than $100$ meV)~\cite{Diallo09,Zhao}.
(II) The observed spin wave excitation is anisotropic in the plane~\cite{Diallo09,Zhao}; this anisotropy has been also found in the paramagnetic (PM) state in the electron-doped system~\cite{Diallo10}.
(III) The wave-vector-averaged intensity of the excitation spectra is lower in the AFM case than that in the PM case below 80 meV~\cite{Lester}.
Most of the experimental data have been analyzed in terms of a local-spin model~\cite{Zhao,Applegate}, except for Ref.~\cite{Diallo09}.

Some other properties of the spin excitations in AFM metals have been illuminated in theoretical work~\cite{Brydon,Kariyado}, but a sufficient understanding of the itinerant AFM phase is not yet provided.
Since understanding the itinerant AFM system is a key to progress in the study of iron pnictide superconductors, we aim to find an appropriate recipe that can describe the basic properties of the AFM phase.
Our recent work~\cite{kaneshita} on modeling the AFM phase of the iron pnictides revealed the weakly ordered character of the ground state by analysis of the optical conductivity.
Based on this success of the ground-state analysis, we attempt here to analyze the spin excitations in the AFM ground state by extending the mean-field approach.

In the next section, we will show our mean-field calculation and the results obtained from those calculations.
In the section after the next, we investigate the spin excitations in an itinerant AFM system by mean-field calculations with a random phase approximation (RPA) within the five-band model that can well describe iron pnictides.
By comparing our results with those of the experiments for the 122 system, we demonstrate how the mean-field model consistently describes the spin-excitation properties, in terms of the presence of the spin wave excitation up to a high energy [characteristics (I)], its anisotropic behavior [characteristics (II)], and the spectral intensity relation between the PM and AFM states [characteristics (III)].
Based on the calculation of the particle-hole excitation spectra, we show that such excitations cause damping of the spin wave excitation, predicting that the spin wave excitation in LaFeAsO (the so-called 1111 system) disappears in a lower energy than in the 122 case.
In the following section, we discuss the difference between the transverse and longitudinal modes of the spin excitations, and also provide a prediction about a charge excitation for a resonant inelastic x-ray scattering (RIXS) experiment.

\section{Mean-field five-band model}
Considering an Fe square lattice (the Fe-Fe bond length $a_0$ set to be unity; the $x$ and $y$ directions, along the nearest Fe-Fe bonds), we start with the five-band mean-field Hamiltonian represented by the ordering vector $\bm{Q}=(\frac{2\pi}{N_Q},0)$ ($N_Q=2$ for AFM, and 1 for PM):
\begin{eqnarray}
&&\hspace{-1cm}H_{MF}=\frac{1}{N_Q}\sum_{\bm{k},\s}\sum_{l,l'}\sum_{\mu,\nu}
\,c_{\bm{k}+l\bm{Q}\, \mu\, \s}^\dagger\, c_{\bm{k}+l'\bm{Q}\, \nu\, \s}
\nonumber\\
&&\hspace{.5cm}\[H_{l}^{(1)}(\bm{k},\s)\,\delta_{l,l'}+
+H_{ll'}^{(2)}(\bm{k},\s)\,\(1-\delta_{l,l'}\)\],
\label{eq:H}
\end{eqnarray}
where $c_{\bm{k}\,\nu\,\s}^\dagger$ creates an electron with wave vector $\bm{k}$ and spin $\s$ at orbital $\mu$.
The diagonal component of $H_{MF}$ is
\begin{eqnarray}
&&\hspace{-.5cm}H_{l}^{(1)}(\bm{k},\s)=
\sum_{\bm{\Delta}} t(\Delta_x,\Delta_y;\mu,\nu)
\,\e^{\i (\bm{k}+l\bm{Q}) \cdot \bm{\Delta}}\
+\epsilon_{\mu}\,\delta_{\mu,\nu},
\end{eqnarray}
where $t(\Delta_x,\Delta_y;\mu,\nu)$ and $\epsilon_{\mu}$ are the tight-binding energies presented in Ref.~\cite{Kuroki}, and $\bm{\Delta}=(\Delta_x,\Delta_y)$.
The off-diagonal component is
\begin{eqnarray}
&&H_{ll'}^{(2)}(\bm{k},\s)
\nonumber\\
&&
\hspace{0.5cm}=
-J\(\sum_{\nu'}\<n_{(l-l')\bm{Q}\, \nu'\nu'\, \s}\>^*
-\<n_{(l-l')\bm{Q} \,\mu\mu\, \s}\>^*\)
\delta_{\mu,\nu}
\nonumber\\
&&\hspace{1.0cm}+J
\(2\<n_{(l-l')\bm{Q} \,\mu\nu\, \s}\>^*
-\<n_{(l-l')\bm{Q} \,\nu\mu\, \s}\>^*\)
(1-\delta_{\mu,\nu})
\nonumber\\
&&\hspace{1.0cm}
-U \<n_{(l-l')\bm{Q} \,\mu\mu\, \s}\>^*,
\end{eqnarray}
where $U$ is the intraorbital Coulomb interaction, $J$ is the Hund coupling, and the pair hopping is set equal to $J$.
Hamiltonian (\ref{eq:H}) is derived from the tight-binding+($U,J$) Hamiltonian represented in Ref.~\cite{kaneshita} by retaining the spin-density-wave order parameter defined as
\begin{eqnarray}
\<n_{l\bm{Q}\,\mu\nu\,\s}\>=\frac{1}{N}\sum_{\bm{k}}
\<c_{\bm{k}+l\bm{Q}\,\mu\, \s}^\dagger\, c_{\bm{k}\,\nu\, \s}\>,
\end{eqnarray}
where $N$ is the number of $\bm{k}$ points in the first Brillouin zone (BZ) of the five-band PM system, and $l\neq0$.

\begin{figure}[t]
\begin{center}
\includegraphics[width = 0.8\linewidth]{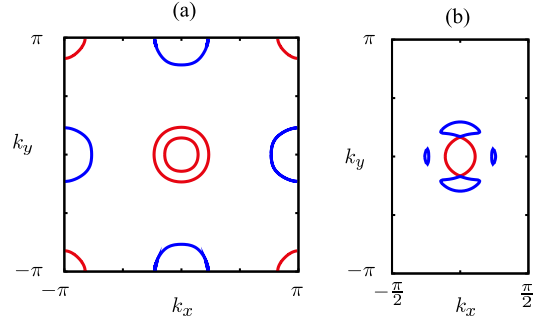}
\caption{(Color online) Fermi surfaces for the PM case (a) and the 122 case (b).
} \label{fig:fermi}
\end{center}
\end{figure}

\begin{figure}[t]
\begin{center}
\includegraphics[width = 0.7\linewidth]{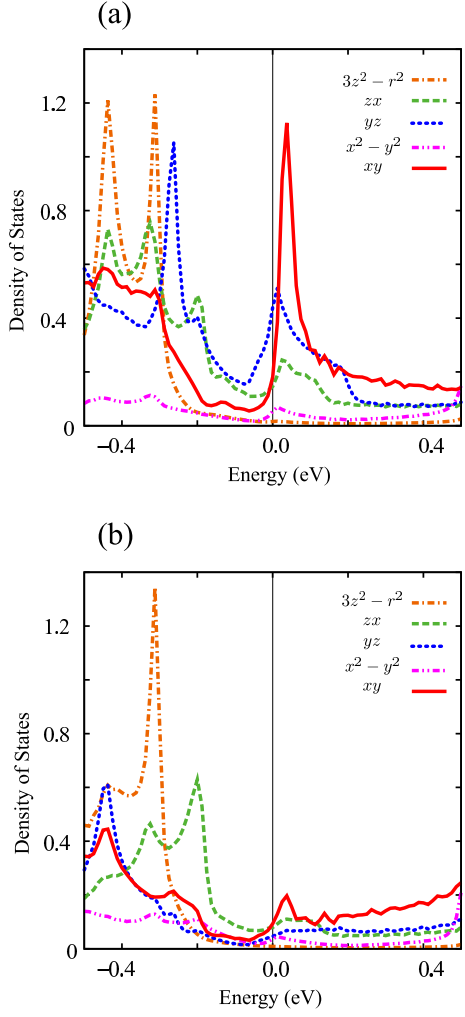}
\caption{(Color online) The orbital-resolved density of states of the majority spin (a) and the minority spin (b).
} \label{fig:dos}
\end{center}
\end{figure}

To obtain the ground state, we solve mean-field equations self-consistently to obtain the quasiparticle state
\begin{eqnarray}
\gamma^{\dagger}_{\bm{k}\,n\,s}=\sum_{\mu,l}
\psi_{\mu,l;n}(\bm{k},\s)
c_{\bm{k}+l\bm{Q}\, \mu \, \s}^\dagger
\end{eqnarray}
with the energy $E_{\bm{k},n,\s}$.
The ground state is represented as a set of quasiparticles distributed according to the Fermi distribution function $f$.
Different parameter sets yield different strength of the order.
We evaluate the order strength from the magnetic moment:
\begin{eqnarray}
M=\sum_{\mu}\<n_{\bm{Q}\,\mu\mu\,\up} -n_{\bm{Q}\,\mu\mu\,\dn}\> \mu_\mathrm{B}.
\end{eqnarray}
The parameter set $U=1.1$ eV and $J=0.2$ eV yields $M=0.4\mu_\mathrm{B}$ corresponding to the 1111 system~\cite{Cruz} (we refer to this as the 1111 model).
The Fermi surface of this model is plotted in Ref.~\cite{kaneshita}.
To simulate the 122 system~\cite{Huang}, we use the 10\% larger $U=1.2$ and $J=0.22$~\cite{Miyake}, which yields $M=0.8\mu_\mathrm{B}$ (the 122 model).
The symmetry-broken Fermi surface of this model is plotted in Fig.~\ref{fig:fermi} together with the symmetric one in the PM case.
We note that Dirac cones appear near $(0,0)$ along the $k_x$ axis as a small electron pocket in the AFM case and that the presence of the Dirac cones affects the transport properties~\cite{Morinari}.
The order strength is also evaluated from the partially opened gap, which is estimated to be $\sim0.2$ eV (122) and $\sim0.1$ eV (1111) from the density of states (for the 1111 model, shown in Ref.~\cite{kaneshita}).
The gap in the single-particle excitation near the $\Gamma$ point opens at around $\bm{k}=(\pm0.2\pi, \pm0.2\pi)$, as can be found in Fig.~\ref{fig:fermi}.
From the orbital-resolved density of states plotted in Fig.~\ref{fig:dos}, it is also found that the $xy$ orbital component is dominant near above the Fermi level.

\section{Dynamical susceptibilities}
To investigate the spin excitations around the mean-field ground state, we calculate the dynamical susceptibility in the spin transverse channel by RPA in Matsubara form,
\begin{eqnarray}
&&\hspace{-.5cm}\chi^{+}_{{\nu\lambda}\atop{\mu\tau}}(\bm{k}_1,\bm{k}_2,\i\tilde{\omega})
= {\chib^{+}}_{{\nu\lambda}\atop{\mu\tau}}(\bm{k}_1,\bm{k}_2,\i\tilde{\omega})
\nonumber\\
&&\hspace{.3cm}- \sum_{\bm{k}'}
\sum_{{{\mu',\nu'}\atop{\lambda',\tau'}}}
{\chib^{+}}_{{\nu\lambda'}\atop{\mu\tau'}}(\bm{k}_1,\bm{k}',\i\tilde{\omega}) V^{++}_{{\lambda'\nu'}\atop{\tau'\mu'}} \chi^{+}_{{\nu'\lambda}\atop{\mu'\tau}}(\bm{k}',\bm{k}_2,\i\tilde{\omega}),
\label{eq:chi+}
\end{eqnarray}
where $\chib$ is the bare susceptibility whose explicit form is given below, the superscript $+$ represents a pair of a down-spin ($\dn$) hole and an up-spin ($\up$) electron, and the nonzero elements of the interaction matrix $V^{++}$ are
\begin{eqnarray}
V^{++}_{{\lambda\nu}\atop{\tau\mu}}
=\left\{
\begin{matrix}
U&  \mbox{for $\lambda=\tau=\mu=\nu$} \\
J& \mbox{for $\lambda=\tau \neq \mu=\nu$}\\
J& \mbox{for $\lambda=\mu \neq \tau=\nu$}\\
U-2J& \mbox{for $\lambda=\nu \neq \mu=\tau$}
\end{matrix}
\right..
\end{eqnarray}

The dynamical susceptibilities of the longitudinal modes are calculated from
\begin{eqnarray}
\begin{pmatrix}
\chi^{\up\up}\\
\chi^{\dn\up}\\
\end{pmatrix}
=
\begin{pmatrix}
\chib^{\up}\\
0
\end{pmatrix}
-
\begin{pmatrix}
\chib^{\up}\,V^{\up\up}&\chib^{\up}\,V^{\up\dn}\\
\chib^{\dn}\,V^{\dn\up}&\chib^{\dn}\,V^{\dn\dn}
\end{pmatrix}
\begin{pmatrix}
\chi^{\up\up}\\
\chi^{\dn\up}
\end{pmatrix},
\end{eqnarray}
where the orbital indices are omitted together with their summations, which are taken in the same manner as in the transverse case, Eq.~(\ref{eq:chi+}).
The nonzero elements of the interaction matrix $V^{\s\s'}$ are
\begin{eqnarray}
V^{\s\s'}_{{\lambda\nu}\atop{\tau\mu}}
=\left\{
\begin{matrix}
U&  \mbox{for $\lambda=\tau=\mu=\nu$} \\
J& \mbox{for $\lambda=\tau \neq \mu=\nu$}\\
U-3J-J\delta_{\s,\s'}& \mbox{for $\lambda=\mu \neq \tau=\nu$}\\
J-(U-2J)\delta_{\s,\s'}& \mbox{for $\lambda=\nu \neq \mu=\tau$}
\end{matrix}
\right..
\end{eqnarray}

The bare susceptibility in Matsubara form is represented with the wave functions and the quasiparticle energies,
\begin{eqnarray}
&&{\chib^{s}}_{{\nu\lambda}\atop{\mu\tau}}(\bm{k}+l_1\bm{Q},\bm{k}+l_2\bm{Q},\i\tilde{\omega})
\nonumber\\
&=&-\frac{1}{N} \sum_{\bm{p_0}}\sum_{n,m}\sum_{l,l'}
\frac{f(E_{\bm{p_0}+\bm{k},n,\s})
-f(E_{\bm{p_0},m,\s'})}
{E_{\bm{p_0}+\bm{k},n,\s}
-E_{\bm{p_0},m,\s'}-\i\tilde{\omega}}
\nonumber\\
&&\hspace{1cm}\times\,
\psi^*_{\nu,\,l_1+l;\,n}(\bm{p}_0+\bm{k},\s)\,
\psi_{\lambda,\,l_2+l';\,n}(\bm{p}_0+\bm{k},\s)
\nonumber\\
&&\hspace{1cm}\times\,
\psi_{\mu,\,l;\,m}(\bm{p}_0,\s')\,
\psi^*_{\tau,\,l+l';\,m}(\bm{p}_0,\s').
\end{eqnarray}
where the set of spins $(\s,\s')$ takes $(\up,\up)$, $(\dn,\dn)$, and $(\up,\dn)$ for $s=\up$, $\dn$, and $+$, respectively.
The summation of $\bm{p}_0$ runs over the reduced magnetic BZ.

We evaluate the imaginary part of the dynamical and the bare susceptibilities
\begin{eqnarray}
\chi''(\bm{k},\omega)
&=&\sum_{\mu,\nu}
\Im{\chi_{{\mu\nu}\atop{\mu\nu}}(\bm{k},\bm{k},\i\tilde{\omega}\rightarrow\omega+\i\eta)}
\end{eqnarray}
to determine the collective and the individual excitations, respectively, where $\eta$ is set to $0.01$ eV.

\section{Spin transverse excitations}
\begin{figure}[t]
\begin{center}
\includegraphics[width = 0.9\linewidth]{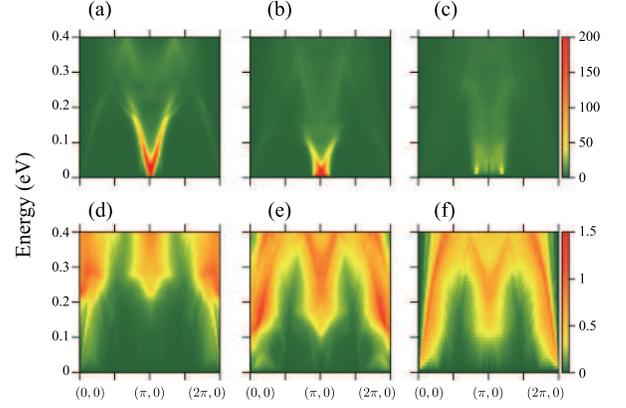}
\caption{(Color online) $({\chi^{+}})''$ [(a)-(c)] and $(\chib^{+})''$ [(d)-(f)] for the $M=0.8\mu_{B}$ [(a), (d)], the $M=0.4\mu_{B}$ [(b), (e)], and the $M=0.0 \mu_{B}$ [(c), (f)] cases.
The color bars are common in (a)-(c) and in (d)-(f), respectively.
Intensities higher than 200 are not taken account of in (a) and (b) for better visibility:
The maximum intensities are $\sim 2800$ (a) and $\sim1000$ (b).
} \label{fig:chix}
\end{center}
\end{figure}

Figures~\ref{fig:chix}(a) and \ref{fig:chix}(b) show the transverse spin-wave excitation spectra in the 122 and 1111 models, respectively.
We find that a collective mode appears at $(\pi, 0)$ and persists up to $\sim0.2$ eV in the 122 model, reproducing the characteristics (I).
Similar spin-wave excitations were obtained within an effective three-band model~\cite{Knolle}.
Above $\sim0.2$ eV, this spin wave excitation is damped; this damping feature has also been noted in Ref.~\cite{Diallo09}.
The 1111 model shows the damping at lower energy ($\sim0.1$ eV).

A possible cause of this damping is particle-hole excitations, as mentioned in Ref.~\cite{Diallo09}; however, there remains a question of why only the high-energy excitations are damped despite the itinerant system.
To investigate this, we plot the particle-hole excitation spectra in Figs.~\ref{fig:chix}(d) and \ref{fig:chix}(e).
The $(\chib^{+})''$ spectra exhibit the strong excitation spectra above 0.2 eV (0.1 eV) in the 122 (1111) model.
This characteristic energy is consistent with the damping feature in $(\chi^{+})''$; therefore, it is clear that the damping is caused by the particle-hole excitations.
Notice that the threshold energies, above which the particle-hole excitations occur, correspond to the partially opened gap estimated from the density of states.

For the PM case, there appears a broad excitation structure around $(\pi,0)$ in  $(\chi^{+})''$ [Fig.~\ref{fig:chix} (c)].
This broad structure is caused by the particle-hole excitations ---the $(\chib^{+})''$ spectra exhibit the gapless feature with some intensities in the low-energy region [Fig.~\ref{fig:chix} (f)].

\begin{figure}[t]
\begin{center}
\includegraphics[width = 0.7\linewidth]{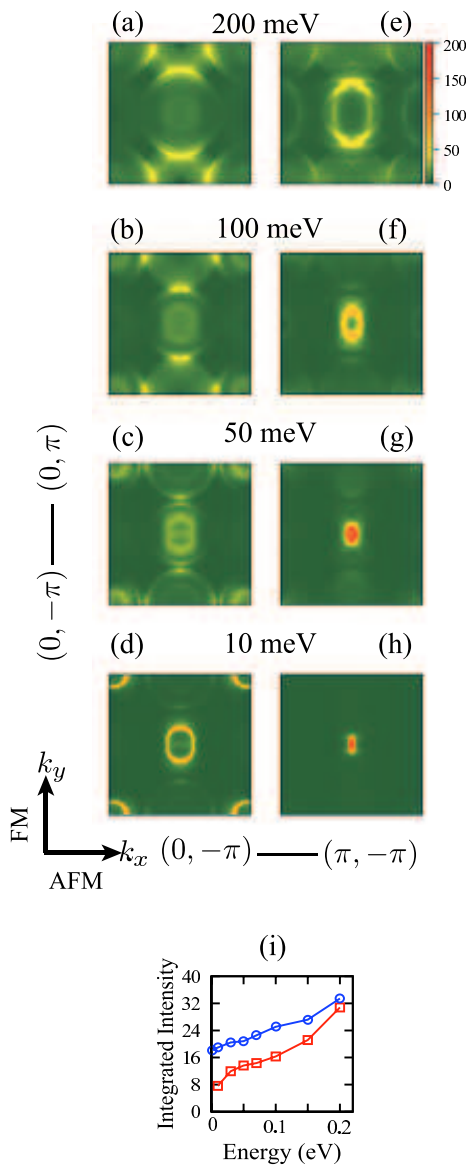}
\caption{(Color online) The intensity map of $(\chi^{+})''(\bm{k},\omega)$ for the PM [(a)-(d)] and the AFM ($M=0.8\mu_B$) [(e)-(h)] cases at the energy $\omega=200$ meV [(a), (e)], 100 meV [(b), (f)], 50 meV [(c), (g)], and 10 meV [(d), (h)].
Intensities higer than 200 are not taken account of in (g) and (h) for better visibility, and the color bar is common in (a)-(h).
Plotted in (i) are the integrated intensities $\frac{1}{4}\sum_{\bm{k}} (\chi^{+})''$ for the PM (circle) and the AFM (square) cases.
} \label{fig:chi-map}
\end{center}
\end{figure}

The energy dependence of the spin-wave cone is plotted in Figs.~\ref{fig:chi-map}(e)-\ref{fig:chi-map}(h).
Our results reproduce the neutron experiments, including the anisotropic structure of the spin-wave cone (II).
In the PM case [Figs.~\ref{fig:chi-map}(a)-\ref{fig:chi-map}(d)], on the other hand, the excitations appear in a wide region around $(\pi,0)$ at each energy.
At low energy [Fig.~\ref{fig:chi-map} (d)], strong intensities lie along the ring around $(\pi,0)$ reflecting the \textit{imperfect} nesting of the Fermi surface
---this excitation ring is anisotropic.
Experimentally observed spectra in the PM state~\cite{Diallo10}, however, show not such a ring structure but a broad spot around $(\pi,0)$.
This broad structure may come from finite-temperature effects that are not included in the calculation.
Strictly speaking, we also need to consider a state-dependent $\eta$ taking into account the different scattering properties in the PM and AFM states.
The problem with the ringlike structure in the PM case requires a further study including such effects to be addressed, and this should be discussed in future work.
Nevertheless, the main point of our interest, the anisotropic feature, is well reproduced.
The anisotropy in the excitation spectra of the AFM and PM states is owing to the anisotropic structure of the bare susceptibility arising from the energy band structure.

Comparing integrated intensity over the momentum space in the AFM and PM cases [Fig.~\ref{fig:chi-map} (i)], we find that the PM state admits more spin transverse excitations than the AFM state.
This result matches the spectral intensity relation (III) between the PM and AFM states.

In our results of the AFM case, no gapped feature is found at $\sim7$ meV, inconsistent with the experiments~\cite{Diallo10,Zhao08,McQueeney,Matan}.
To reproduce the spin-gap feature, we presumably need to take into account effects not included in our calculation, such as the single-ion anisotropy~\cite{Zhao08}.

\begin{figure}[t]
\begin{center}
\includegraphics[width = 0.9\linewidth]{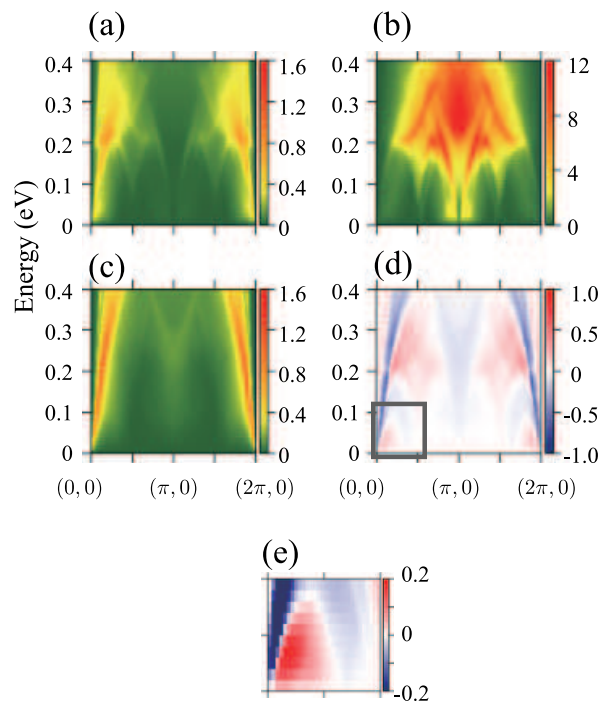}
\caption{(Color online) (a) $({\chi^{n}})''$ for the 122 case, (b) $({\chi^{z}})''$ for the 122 case, and (c) $({\chi^{n}})''$ for the PM case.
(d) The difference between the AFM and PM cases [(c) subtracted from (a)].
(e) Detail of the squared region in (d).
} \label{fig:chil}
\end{center}
\end{figure}

\section{Spin longitudinal and charge excitations}
Now we analyze the spin longitudinal and charge excitation modes.
The excitation spectra for the modes [$\chi^n=\chi^{\up\up}+\chi^{\dn\dn}$, $\chi^z=\frac{1}{2} (\chi^{\up\up}-\chi^{\dn\dn})$] in the 122 case are plotted in Figs.~\ref{fig:chil}(a) and \ref{fig:chil}(b).
The intensity of these modes, in contrast to the spin transverse mode, is weak in the low-energy region; this means that the particle-hole excitations without a spin flip inside the gap is strong enough to damp the low-energy excitations of these modes.

To investigate the charge dynamics, we propose a RIXS experiment on these materials, which can measure the momentum-resolved charge excitations.
For this purpose, we examine the details of the charge excitation mode from a theoretical point of view.
Compared to the PM state [Fig.~\ref{fig:chil} (c)], which shows a rodlike structure rising from $(0,0)$, the excitations in the AFM state are rather broad.
This difference arises from the presence of the magnetic order with which the charge collective excitations cost more energy accompanied by the spin excitations of the longitudinal mode.
The difference in the spectra becomes clearer in the plot of the subtraction [Fig.~\ref{fig:chil} (d)], where the rodlike excitation structure in the PM state becomes narrow and enhanced; in RIXS, this change should be observable ---for example, the change comparable to $40$\% of the peak intensity is found at 0.4 eV.

We also find the difference in the low-energy excitations around $(0,0)$ [Fig.~\ref{fig:chil} (e)].
Since the spectra of the AFM case shows a broad structure there, low-energy charge fluctuations can occur with various wave vectors away from $(0,0)$.
These low-energy charge fluctuations may be related to nematic charge structures observed by spectroscopic imaging-scanning tunnel microscopy~\cite{Chuang}, where an eight-site periodic structure represented by the wave vector $(\pi/4,0)$ is observed.

\section{Conclutions}
In summary, we have investigated the spin excitations in an itinerant AFM system by mean-field calculations within an RPA in a five-band model.
Our results have reproduced the characteristics (I)-(III) observed in experiments for the 122 system.
In the spin-wave excitation, the states outside the partially opened gap plays an important role.
Particle-hole excitations across the gap with a spin flip cause the damping of the spin-wave excitations above the threshold energy corresponding to the partially opened gap.
On the other hand, such excitations within the gap are too weak to damp the collective excitation.
We predict that the spin-wave excitation in the 1111 system disappears at a lower energy than in the 122 case since the magnetic moment ---directly related to the partially opened gap--- is smaller in the 1111 system.
So far, the inelastic neutron scattering in the 1111 system has been performed only on a powder sample~\cite{Ishikado}; the experiment on a single crystal is desired.

In contrast to the spin transverse mode, the charge and spin longitudinal modes are weak even in the low energy region, because the particle-hole excitations arising from the states inside the gap damp these excitations.
From the analysis of the charge excitation, we provide a prediction for a RIXS experiment.
A rodlike structure should be observed in the difference of the excitation spectra for the AFM and PM cases.
In addition, we have found that the magnetic order causes the charge fluctuations.
These charge fluctuations lying away from $(0,0)$ may be related to the nematic charge structure observed recently, and this structure would not exist in the PM state, where such charge fluctuations are found to be weak.
This structure formation should involve the $xy$ orbital component, which is dominant in the states in the energy range 0-50 meV.

In conclusion, all the above features of the spin and charge dynamics are ruled by the magnetic order, and the characteristic energies for damping and the strength of the charge fluctuations are expected to scale with the magnetic order strength, i.e., the magnetic moment.

\section{Acknowledgments}
We thank T. Morinari and A. R. Bishop for useful comments.
This work was supported by the Grant-in-Aid for Scientific Research from the Ministry of Education, Culture, Sports, Science and Technology of Japan; the Global COE Program ``The Next Generation of Physics, Spun from University and Emergence"; the Next Generation Supercomputing
Project of Nanoscience Program; and Yukawa Institutional Program for Quark-Hadron Science at YITP.
Numerical computation in this work was carried out at the Yukawa Institute Computer Facility.
A part of the work done by E. K. is supported by Yukawa Memorial Foundation.


\end{document}